# On the Overestimation of Efficiency in Relativistic Electron Scattering

Grant Brassem[1], Christian Viernes[1], Germán Sciaini[1,*]

[1]The Ultrafast Electron Imaging Laboratory, Department of Chemistry and Waterloo Institute for Nanotechnology, University of Waterloo, 200 University Ave. W., Waterloo, N2L 3G1, Ontario, Canada.

Address correspondence to: Germán Sciaini; gsciaini@uwaterloo.ca.

**Recent reviews in ultrafast electron diffraction (UED) have claimed that relativistic electrons exhibit enhanced elastic scattering efficiency, frequently quantified as a $\gamma^2$ increase in the differential cross section. These claims, however, originate from angular-domain analyses that overlook the compression of scattering angles $\theta$ with increasing electron energy, leading to an apparent—but artificial—enhancement. In this work, we recast the problem in momentum-transfer space $q$, where scattering is accurately accounted for. This transformation eliminates the angular compression artefact and reveals that high-energy scaling follows a simple $\beta^{-2}$ dependence, with no intrinsic relativistic gain. We demonstrate this by directly integrating relativistic differential elastic-scattering cross sections from ELSEPA and by applying a straightforward transformation of the well-known Mott–Massey formalism into $q$-space. The results are general, with calculations performed for elements from carbon to gold and for energies between 50 keV and 5000 keV. They reproduce the long-established trend in total elastic scattering cross sections, in which scattering strength decreases with increasing electron kinetic energy. Practically, at energies above roughly 50 keV, scattering is already dominated by the forward direction, and most of the scattered intensity falls within the acceptance range of typical UED detectors. These findings correct a widespread misconception in the UED literature and provide a more accurate and intuitive framework for interpreting and optimizing high-energy electron scattering experiments.**

# INTRODUCTION

Ultrafast electron diffraction[1] (UED) has emerged as a powerful method for probing structural dynamics in matter with femtosecond temporal resolution and atomic-scale spatial precision. By using short pulses of electrons to interrogate a sample, UED captures "snapshots" of atoms in motion during processes such as phase transitions, chemical reactions, or lattice vibrations. Electrons interact strongly with matter, providing large scattering cross sections compared to X-rays and enabling high sensitivity to light elements as well as the use of thin films or nanostructured samples. At typical acceleration voltages, their sub-ångström de Broglie wavelengths allow for the resolution of molecular structures with high accuracy.

Recent technological developments—most notably the advent of relativistic, radio-frequency–driven MeV electron sources[2]—have broadened the operating range of UED, reducing space-charge broadening, increasing penetration depth, and, in certain regimes, providing relatively better temporal resolution through higher extraction fields. However, these developments have also led to claims that relativistic electrons inherently possess greater elastic scattering efficiency, often cited as scaling with $\gamma^2$ according to the formalism introduced by Mott and Massey[3]. Such assertions, if true, would have significant implications for experiment design and optimization in high-energy UED settings.

Yet, as we show in this work, these statements are based on angular-domain interpretations that neglect the compression of scattering angles with increasing kinetic energy. By reframing the problem in momentum-transfer space $q$, we reveal that there is no intrinsic relativistic gain in scattering efficiency, and that the scaling follows a simple $\beta^{-2}$ dependence—consistent with long-established trends in total elastic-scattering cross sections.

# RESULTS AND DISCUSSION

## Direct Integration of Differential Electron Elastic-Scattering Cross Sections

In this study, we examine the impact of relativistic effects on the differential and total elastic-scattering cross sections in electron diffraction. The electron–atom scattering cross section determines the probability and angular distribution of scattering events, directly shaping the observed diffraction intensities in UED experiments. To this end, we use ELSEPA[4,5] to perform Dirac partial-wave calculations for elements ranging from carbon (C) to gold (Au) over energies from 50 to 5000 keV. The differential cross section is usually defined as area per unit solid angle $\Omega$ (i.e., $\frac{d\sigma}{d\Omega}$). Although most studies analyze high-energy scattering solely in terms of angular deflection $\theta$, momentum transfer provides a more physically meaningful metric because it directly reflects the change in the electron's momentum, is independent of the beam energy-dependent compression of scattering angles and allows for consistent comparison across different energies and scattering geometries.

The momentum transfer $q$ is defined as:

$$q = \frac{4\pi \sin(\theta/2)}{\lambda} \qquad (1)$$

where $\lambda$ is the de Broglie wavelength given by:

$$\lambda = \frac{h}{\gamma m_o \beta c} \qquad (2)$$

Here, $h$ is the Max Planck constant, $m_o$ is the mass of the electron at rest, $c$ is the speed of light in the vacuum, $\beta = v/c$ is the ratio between the electron's speed $v$ and $c$, and $\gamma = (1 - \beta^2)^{-1/2}$ is known as the Lorentz factor.

We now calculate the values of effective electron elastic-scattering cross sections, $\sigma_{\text{eff}}$, within the desired maximum momentum-transfer limit $q_{\max}$ through integration:

$$\sigma_{\text{eff}} = 2\pi \int_0^{\theta_{\max}} \frac{d\sigma}{d\Omega} \sin\theta \, d\theta \qquad (3)$$

provided that $\theta_{\max} = 2 \sin^{-1}(q_{\max} \lambda/4\pi)$ and $d\Omega = 2\pi \sin\theta \, d\theta$ as defined in ELSEPA; where the solid angle has been azimuthally integrated, and therefore $\frac{d\sigma}{d\Omega}$ is a function of $\theta$.

Thus, we have everything needed to calculate the effective $\sigma_{\text{eff}}$ and total $\sigma_{\text{total}}$ electron elastic-scattering cross sections. The latter values are also provided by ELSEPA and serve as checkpoints, confirming the implemented definition of $d\Omega$ in our integration.

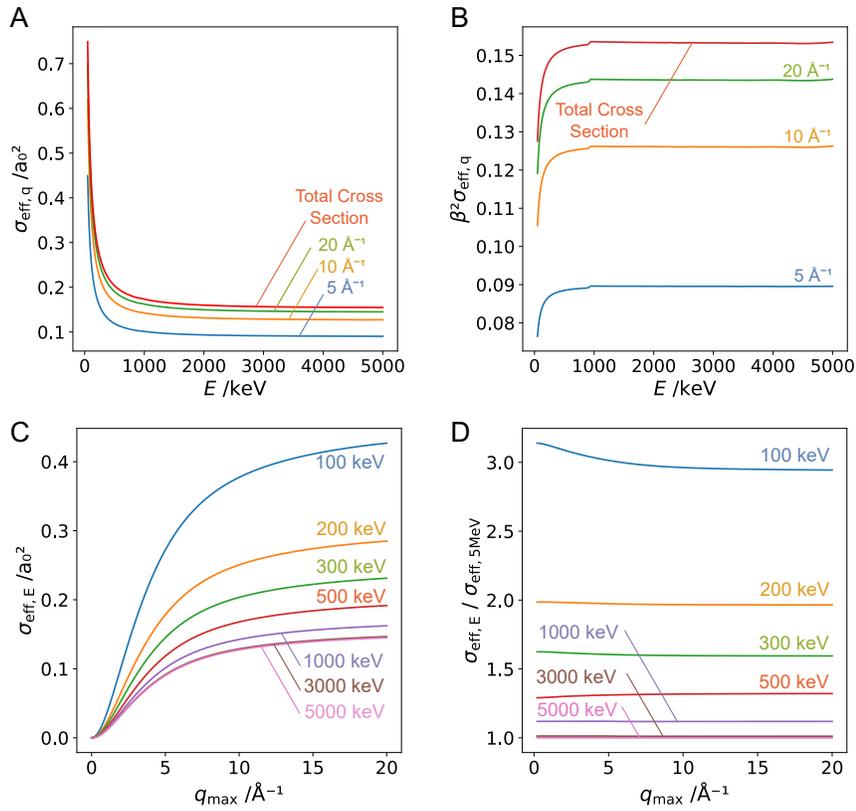

**Fig. 1.** Effective $\sigma_{eff}$ and total elastic cross sections for Ag. A) $\sigma_{eff}$ values calculated as a function of electron energy $E$ for maximum momentum-transfer limits $q_{max}$ = 5, 10 and 20 Å$^{-1}$. B) $\beta^2 \sigma_{eff}$ as a function of $E$ for the same $q_{max}$ values. The discontinuity near 1000 keV arises from ELSEPA switching to a faster but less accurate model at high energies; this effect is also visible, though less pronounced, in panel A. C) $\sigma_{eff}$ as a function $q_{max}$ for $E$ = 100, 200, 300, 500, 1000, 3000 and 5000 keV. D) Ratios $\sigma_{eff,E}/\sigma_{eff,5\,MeV}$ as a function of $q_{max}$ for the same $E$ values, with $E$ = 5000 keV taken as the reference.

Figure 1 presents $\sigma_{eff}$ as a function of electron energy $E$ and $q_{max}$. As evident from panels A and C, the overall shapes of the traces are preserved, with no observable crossovers in either the $E$ or $q$ domains. This is further illustrated in panels B and D. In panel B, multiplying $\sigma_{eff}$ by $\beta^2$ yields approximate plateau values, indicating that the dependence of both $\sigma_{eff}$ and $\sigma_{total}$ on $E$ follows a $\beta^{-2}$ scaling—a point that will be discussed later. Panel D shows the ratios of $\sigma_{eff}$ at various $E$ values to $\sigma_{eff}$ at $E$ = 5000 keV (taken as a reference) as a function of $q_{max}$. Figure 1D demonstrates that the main differences arise primarily from simple proportionality constants. The calculations further indicate that there is no relativistic gain; on the contrary, the results are consistent with expectations from total electron elastic-scattering cross sections.

**Revisiting Mott and Massey's Formalism**

To gain a better understanding of the dominating factors in the observed behaviors, we turn to the well-known relativistic formula for the differential electron elastic-scattering cross section, $\frac{d\sigma}{d\Omega}$, introduced by Mott and Massey[3],

$$\frac{d\sigma}{d\Omega} = \frac{1 - \beta^2 \sin^2(\theta/2)}{1 - \beta^2} |f(q)|^2 \qquad (4)$$

Where $f(q)$ is the *sine* Fourier transform of the Coulomb potential $V(r)$,

$$f(q) = -\frac{2 m_o}{\hbar^2} \int_0^\infty \frac{\sin(qr)}{qr} V(r) r^2 \, dr \qquad (5)$$

By assuming that the scattering angle $\theta$ is relatively small, we can approximate in **Eq. 4** that $1 \gg \beta^2 \sin^2(\theta/2)$, which leads to:

$$\frac{d\sigma}{d\Omega} \simeq \frac{1}{1 - \beta^2} |f(q)|^2 = \gamma^2 |f(q)|^2 \qquad (6)$$

The erroneous use of **Eq. 6** has led some in the UED community to conclude that relativistic electrons are inherently more efficient that their sub-relativistic counterparts. Lee *et al.*[6] implemented a partial correction and reported that the scattering amplitude for electrons at 4 MeV is approximately 7.5 times greater than for electrons at 100 keV, corresponding to a scaling factor of about $\gamma$. Similarly, the review by Filippetto *et al.*[7] notes that the differential cross section versus momentum transfer increases proportionally to $\gamma^2$, essentially following the scaling derived by Mott and Massey[3]. Both reviews cite the work of Zhu *et al.*,[8] who were the first to emphasize that higher electron beam energies lead to larger differential elastic-scattering cross sections. These claims are misleading since they refer to $\frac{d\sigma}{d\Omega}$ or include only partial corrections.

A more rigorous analysis, presented here, transforms the differential cross sections into the appropriate $q$-frame of reference, thereby eliminating the apparent high-energy "gain" by accounting for the angle-compression effect with the change in de Broglie wavelength.

By applying the chain rule:

$$\frac{d\sigma}{dq} = \frac{\partial \sigma}{\partial \Omega} \frac{\partial \Omega}{\partial q} = \frac{\partial \sigma}{\partial \Omega} \frac{\partial \Omega}{\partial \theta} \frac{\partial \theta}{\partial q} \qquad (7)$$

and using the expression for $d\Omega$ given above, we have:

$$\left(\frac{\partial \Omega}{\partial \theta}\right)_\lambda = 2\pi \sin\theta \quad (8)$$

and from differentiating **Eq. 1**:

$$\left(\frac{\partial \theta}{\partial q}\right)_\lambda = \frac{\lambda}{2\pi \cos(\theta/2)} \quad (9)$$

Combining these results gives:

$$\frac{d\sigma}{dq} = \frac{\partial \sigma}{\partial \Omega} 2\pi \sin\theta \frac{\lambda}{2\pi \cos(\theta/2)} = \frac{\sin\theta}{\cos(\theta/2)} \lambda \gamma^2 |f(q)|^2$$

$$\frac{d\sigma}{dq} = 2 \sin(\theta/2) \lambda \gamma^2 |f(q)|^2 \quad (10)$$

and given that, from **Eq. 1**, $\sin(\theta/2) = \frac{q\lambda}{4\pi}$, we obtain:

$$\frac{d\sigma}{dq} = \frac{q\lambda^2}{2\pi} \gamma^2 |f(q)|^2 \quad (12)$$

which, when combined with **Eq. 2** for the de Broglie relationship, becomes:

$$\frac{d\sigma}{dq} = \frac{\hbar h q}{(m_0 \beta c)^2} |f(q)|^2 \quad (13)$$

This shows that the dependence of the differential electron elastic-scattering cross section in the $q$−domain follows a simple $\beta^{-2}$ scaling, consistent with the behavior observed in **Fig. 1B**. Moreover,

$$\frac{\sigma_{\text{eff},E}}{\sigma_{\text{eff},5\text{MeV}}} \approx \left(\frac{\beta_{5MeV}}{\beta_E}\right)^2 \quad (14)$$

yielding $\frac{\sigma_{\text{eff},E}}{\sigma_{\text{eff},5\text{MeV}}}$ ratios of 3.31, 2.06, 1.65, 1.34, 1.13, and 1.02 for $E = 100, 200, 300, 500, 1000$, and $3000$ keV, respectively—values that approximately match the offsets observed in **Fig. 1D**.

This analysis demonstrates that the apparent advantage of relativistic electrons in producing larger cross sections vanishes under rigorous treatment. The generalization of these results is illustrated in **Fig. 2**, which summarizes $\frac{\sigma_{\text{eff},E}}{\sigma_{\text{eff},5\text{MeV}}}$ as a function of $q_{\max}$ for several elements from C to Au.

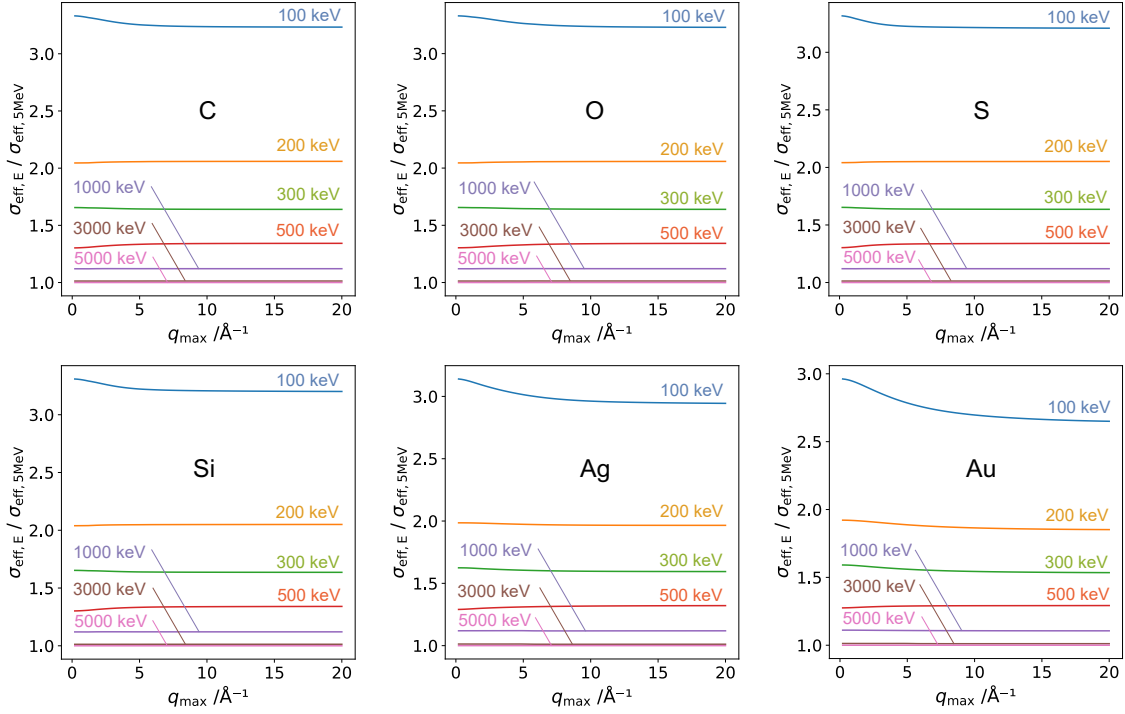

**Fig. 2.** Ratios $\frac{\sigma_{\text{eff},E}}{\sigma_{\text{eff},5\text{MeV}}}$ as of $q_{\text{max}}$ for various elements at $E = 100, 200, 300, 500, 1000, 3000$ and $5000$ keV, with $E = 5000$ keV taken as the reference. The name of each element is indicated in its corresponding panel.

The deviations from the $\beta^{-2}$ scaling seen in **Fig. 1B** at lower $E$—also evident in **Fig. 2**, where the **Eq. 14** proportions deviate more as the atomic number, $Z$, increases—arise from the angular dependence of elastic scattering: lowering the electron energy increases the probability of wide-angle (including backscattering) events, and at fixed energy a higher $Z$ likewise boosts wide-angle scattering. This behavior is consistent with standard scattering physics. However, even for a heavy element like Au, the resulting efficiency loss is rather modest—$\approx (3.3 - 2.7)/3.3 \times 100 = 18\%$ at 100 keV—and is much smaller at 300 keV.

## CONCLUSION

In our view, the main advantage of MeV-electrons is the mitigation of space-charge effects due to relativistic kinematics. This becomes increasingly important for achieving higher spatial resolution in high-brightness ultrafast electron microscopy, where magnetic lens crossovers can otherwise be detrimental.

For UED, the situation is different. In current instruments, diffraction from single crystals is largely limited by sample mosaicity and beam imperfections, which broaden reciprocal-lattice rods (relrods)—an effect that grows with $q$. This behaviour is accounted for in advanced analysis software such as GARDFIELD[9]. As a result, the flatter Ewald sphere at higher energies does not provide a consistent experimental advantage. This is illustrated by UED experiments on the nearly commensurate charge-density-wave (CDW) phase of 1T-TaS$_2$, conducted across electron energies from 30 keV to 3 MeV[8,10–12]. These measurements reach high-$q$ values, with first-order CDW reflections clearly visible in diffraction patterns recorded along the [0,0,1] zone axis—even though $q_{CDW}$ is out of plane. By contrast, in conventional electron microscopes—where the beam is closer to a plane wave and the illuminated areas are much smaller—these reflections are typically not visible[10,13,14].

Our analysis indicates that ≈300 keV provides a practical balance between penetration depth and scattering efficiency, with realistic all-electrostatic electron source designs capable of ≲30-fs temporal resolution[15]. Accordingly, a dedicated 300-kV UED facility operating within WaterFEL[16] will be developed to provide users with access to these performance regimes.

**Acknowledgments:** The authors gratefully acknowledge funding support from the Natural Sciences and Engineering Council of Canada (NSERC) [CREATE 565360] and the Deutsche Forschungsgemeinschaft (DFG, German Research Foundation) [IRTG 2803-461605777]. We also acknowledge funding support from the Canada Foundation for Innovation and the Ontario Research Fund.

**Author contributions:** G. Sciaini conceived the idea and directed the work. G. Brassem conducted the calculations with assistance from C. Viernes. G. Brassem and G. Sciaini wrote the manuscript.

**Competing interests:** The authors declare that there is no conflict of interest regarding the publication of this article.

## DATA AVAILABILITY

A Python program named "Scattering_Efficiency" has been developed and is publicly available through our group repository: https://github.com/UeIL-Waterloo. This tool allows readers to generate any desired plot.